\documentclass[twocolumn,prl,superscriptaddress,showpacs]{revtex4}

\usepackage{graphicx}
\usepackage{amssymb}
\usepackage{amsfonts}

\DeclareGraphicsExtensions{.eps}

\begin{document}

\title{Polariton Squeezing in Semiconductor Microcavities}

\date{\today}

\author{J.Ph. Karr}
\affiliation{Laboratoire Kastler Brossel, Universit\'{e} Paris 6, Ecole
Normale Sup\'{e}rieure et CNRS,\\
UPMC Case 74, 4 place Jussieu, 75252 Paris Cedex 05, France}

\author{A. Baas}
\affiliation{Laboratoire Kastler Brossel, Universit\'{e} Paris 6, Ecole
Normale Sup\'{e}rieure et CNRS,\\
UPMC Case 74, 4 place Jussieu, 75252 Paris Cedex 05, France}

\author{R. Houdr\'{e}}
\affiliation{Institut de Micro et Opto\'{e}lectronique, Ecole Polytechnique F\'{e}d\'{e}rale de Lausanne, Lausanne, CH
1015 Switzerland}

\author{E. Giacobino}
\affiliation{Laboratoire Kastler Brossel, Universit\'{e} Paris 6, Ecole
Normale Sup\'{e}rieure et CNRS,\\
UPMC Case 74, 4 place Jussieu, 75252 Paris Cedex 05, France}

\begin{abstract}
We report squeezed polariton generation using parametric polariton four-wave mixing in semiconductor microcavities in the
strong coupling regime. The geometry of the experiment corresponds to degenerate four-wave mixing, which gives rise to a
bistability threshold. Spatial effects in the nonlinear regime are evidenced, and spatial filtering is required in order
to optimize the measured squeezing. By measuring the noise of the outgoing light, we infer a 9 percent squeezing on the
polariton field close to the bistability turning point.
\end{abstract}


\maketitle

Cavity polaritons are two-dimensional mixed states of light and matter, obtained in high finesse semiconductor
microcavities in the strong coupling regime \cite{weisbuch}. They present a sharply distorted dispersion relation
\cite{houdre} and novel nonlinear features possessed neither by photons nor by excitons alone. Coulomb interaction between
excitons gives rise to an effective nonlinear interaction between polaritons \cite{ciuti}. Strongly nonlinear emission has
been observed under both resonant \cite{magic} and nonresonant \cite{nonres} pumping.

Nonlinear emission under resonant pumping is usually studied by exciting the microcavity at the so-called "magic angle"
fulfilling both energy and in-plane momentum conservation for the conversion of two pump polaritons with wave vector
$\mathbf{k_{p}}$ into a pair of signal and idler polaritons with wave vector $\mathbf{0}$ and $\mathbf{2k_{p}}$,
respectively (polariton four-wave mixing). However, a pronounced nonlinear behavior can also be obtained by exciting the
microcavity at normal incidence \cite{prl,academie,karr,dunbar}. In this configuration, the nonlinear mechanism is
quasi-degenerate four-wave mixing between polaritons with $\mathbf{k \simeq 0}$. The coherent nature of the nonlinear
process was demonstrated by observing the phase dependence of the emission \cite{prl}.

In this paper, we demonstrate that this coherent nonlinear mechanism allows one to beat thermal effects, often dominant in
the noise of semiconductor structures, and to enter into the quantum regime, that is to generate squeezed polaritons.
While light squeezing has been obtained with a number of nonlinear media, including semiconductors \cite{fox}, the
presence of strong coupling inside the microcavity allows to produce a new type of squeezing involving a part-light,
part-matter field, the polariton. Polariton squeezing opens the way to the use of semiconductor microcavities in the field
of quantum information, for the processing of quantum variables.

As a general rule, quasi-degenerate four-wave mixing is well suited to generate squeezing. The effective Hamiltonian for
the lower polariton in the case of resonant excitation at normal incidence is

\begin{equation}
H=E_{LP}(0)p_{0}^{\dag} p_{0} + \frac{1}{2} X_{0}^{4} V^{eff}p_{0}^{\dag} p_{0}^{\dag} p_{0}p_{0} \label{Hami}
\end{equation}

where $p_{0}$, $p_{0}^{\dag}$ are the annihilation and creation operators for a lower branch polariton with in-plane wave
vector $\mathbf{k}=0$, $E_{LP}(0)$ is the lower polariton energy at $\mathbf{k}=0$ and
$V^{eff}=6e^{2}a_{exc}/\epsilon_{0}A$ is the effective potential due to exciton-exciton interaction, where $a_{exc}$ is
the exciton two-dimensional Bohr radius, $\epsilon_{0}$ is the dielectric constant of the quantum well and $A$ is the
quantization area. $X_{0}$ is the Hopfield coefficient \cite{hopfield} representing the exciton content of the polariton.
In view of the low excitation level, exciton-exciton interaction can be treated as a perturbation compared to the normal
mode coupling, allowing to ignore the effect of the upper polariton. Other nonlinear effects \cite{khitrova} such as
exciton bleaching \cite{houdre2} or carrier quantum correlations \cite{ell} are effective for much higher excitation
densities than the ones considered here. We have checked the presence of two peaks in reflectivity for all the used laser
intensities, and measured excitation induced shifts which were always small compared to the vacuum Rabi splitting
\cite{prl}.

This Hamiltonian is analogous to that of an optical field in a cavity containing a Kerr medium. It is well known that very
good squeezing is expected for the optical field in this case \cite{collett}. In the same way, squeezing of the polariton
field has been predicted in semiconductor microcavities \cite{messin,academie,quattropani}. In this paper, we report the
first observation of polariton squeezing. We make measurements on the light going out of the cavity, i.e. the photon part
of the polariton field, which allows to obtain the amount of polariton squeezing inside the cavity.

Unlike experiments using transparent optical Kerr media, excitation in a microcavity is resonant both with the cavity and
with the nonlinear medium. As a consequence, the squeezing rate is expected to be reduced due to the excess noise added by
the polariton photoluminescence. It is important to evaluate the amount of excess noise in order to assess the feasibility
of squeezing. In the framework of the input-output formalism \cite{entreesortie}, the Heisenberg-Langevin equation for the
lower polariton operator $p_{0}$ can be deduced from the Hamiltonian (\ref{Hami}):

\begin{eqnarray}
\frac{dp_{0}}{dt} &=& -(\gamma_{0} + i \delta_{0})p_{0} - i X_{0}^{4} \alpha p_{0}^{\dag} p_{0}p_{0} \nonumber \\
&-& C_{0}\sqrt{2\gamma_{cav}}A^{in}+X_{0}\sqrt{2\gamma_{exc}}B^{in} \label{dyna}
\end{eqnarray}

In this equation $\delta_{0}=E_{LP}(0)/\hbar-\omega_{L}$ is the detuning from the polariton of the excitation laser of
frequency $\omega_{L}$~; $\gamma_{0}$, $\gamma_{cav}$ and $\gamma_{exc}$ are the linewidths of the polariton, cavity and
exciton. $A^{in}$ ($B^{in}$) is the input term due to the photonic (excitonic) part of the polariton, $C_{0}$ and $X_{0}$
are the Hopfield coefficients and $\alpha=V^{eff}/\hbar$.

\begin{figure}[h]
\centerline{\includegraphics[clip=,width=8.1cm]{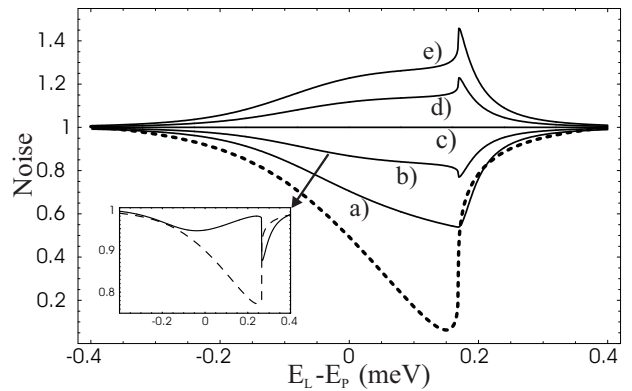}} \caption{Solid line: minimal noise S of the reflected light
(normalized to standard quantum noise) at zero noise frequency versus the laser-polariton energy detuning $E_{L}-E_{P}$,
for various values of the excess noise parameter: (a) $\beta$=0, (b) $\beta$=0.5 $\beta_{c}$, (c) $\beta$=$\beta_{c}$, (d)
$\beta$=1.5 $\beta_{c}$, (e) $\beta$=2 $\beta_{c}$. Dashed line: reflectivity. Other parameters: cavity-exciton detuning
$\delta$=0, input intensity $I_{1}^{in}$ is equal to the bistability threshold intensity. Curve (c) is indistinguishable
from the horizontal axis $S_{min}$=1. Inset: minimal noise obtained with $\beta$=0.5 $\beta_{c}$, and including the effect
of collisional broadening. The exciton linewidth is $\gamma_{exc} (\mu eV) = 75 + 5 n_{b}$ where $n_{b}$ is the exciton
mean number, accounting for the measured dependence of the polariton linewidth on the density of excitation.} \label{fig1}
\end{figure}

Using this equation, the mean values of the polariton field inside the microcavity and of the optical field outside can
easily be calculated. Distortion of the reflectivity curve as well as optical bistability are obtained, as shown by the
dashed line in Fig.~\ref{fig1}. Evidence for a bistable behavior is given in Fig.~\ref{fig3}. This new effect will be
described in more detail in a forthcoming publication. Bistability has been predicted already in semiconductor
microcavities \cite{tredicucci,chen}, however in different conditions.

We now turn to the study of fluctuations. The input term for the light field $A^{in}$ is the driving laser field, which
has fluctuations equal to the vacuum noise. The input term for the exciton field $B^{in}$ comprises only noise since there
is no direct excitation of the exciton field. Let us stress that the presence of relaxation processes for the exciton
implies the existence of fluctuations, due to fluctuation-dissipation theorem. These input fluctuations are at least the
vacuum noise. In addition there is thermal noise coming from phonon interaction. Assuming that the noise is due to an
exciton reservoir which is populated by collisions of the pump polaritons with acoustic phonons, we find that the
correlation function of the input noise is $<B^{in \dag}(t) B^{in}(t+\tau)>=\beta n_{b}\delta(\tau)$, where $n_{b}$ is the
population of the $\mathbf{k}=0$ exciton mode and $\beta$ is the excitation rate of the reservoir \cite{karr}.

Using these assumptions for the polariton excess noise input, we have calculated from equation (\ref{dyna}) the noise
properties of the light going out of the microcavity. This noise is made of a superposition of quadrature components with
phases distributed over 2$\pi$. In the linear regime this emission has no phase dependence, i.e. all the quadrature
components have the same intensities. As shown previously \cite{prl}, in the presence of nonlinear polariton interaction
the noise exhibits a phase dependence, i.e. some quadrature components are amplified while some other quadrature
components are deamplified. Using a linearized version (to lowest order) of equation (\ref{dyna}) and of its hermitian
conjugate allows to predict the behavior of all the noise quadratures. Here, we will concentrate on the quadrature that
exhibits the minimal noise.

Fig.~\ref{fig1} shows the minimal noise as a function of the laser detuning, for various excess noise parameters, zero
cavity-exciton detuning and an input intensity corresponding to the bistability threshold. One obtains squeezing provided
the parameter $\beta$ is smaller than a critical value $\beta_{c}=\alpha/2 \gamma_{exc}$ \cite{karr}. The best squeezing
is obtained in the vicinity of the bistability turning point.

The value of $\beta$ can be evaluated experimentally by measuring the noise at very low excitation density. In such
conditions the nonlinear term in equation (\ref{dyna}) can be neglected and the reflected field has thermal excess noise
increasing linearly with the excitation intensity. We verified this and deduced the value of $\beta$ from the slope of the
excess noise vs the intensity. It was found to be smaller than $\beta_{c}$ for all detunings. This means that the level of
excess noise should be low enough to get squeezing. At this point, this treatment ignores other noise sources like the
exciton-exciton interaction itself that is at the origin of the coherent nonlinear effect but also causes incoherent
effects such as broadening and fluctuations.

This effect can in principle be calculated from the Hamiltonian (\ref{Hami}) \cite{gardiner}. It is however quite
difficult to evaluate by this method. We have estimated it phenomenologically by using the fluctuation-dissipation
theorem. For this, we replace $\gamma_{exc}$ in eq. (\ref{dyna}) by its value in the presence of collisional broadening,
that is, we take into account the incoherent effect of exciton-exciton interaction, the coherent part of which is
accounted for by the second term on the right hand side of eq. (\ref{dyna}). As a result, additional dissipation is
included by means of the first term and additional fluctuations are included by means of the last term. Using the measured
value of the broadening, we obtain the curve shown in the inset of Fig.~\ref{fig1}, which shows an overall reduced
squeezing, as compared to the squeezing predicted without broadening for the same value of $\beta$=0.5 $\beta_{c}$. It can
also be noted that, in this case no squeezing is predicted around the minimum of the reflectivity curve. This can be
understood from the fact that in this region, the driving laser light intensity is high, and so is the polariton number,
which leads to an increased collision rate. On the contrary, right below the bistability threshold, the collision rate is
reduced, while the coherent nonlinear interaction is still high enough to allow for squeezing.

In the experiments presented below, we show how this squeezing can be evidenced.

\begin{figure}[h]
\centerline{\includegraphics[clip=,width=8.1cm]{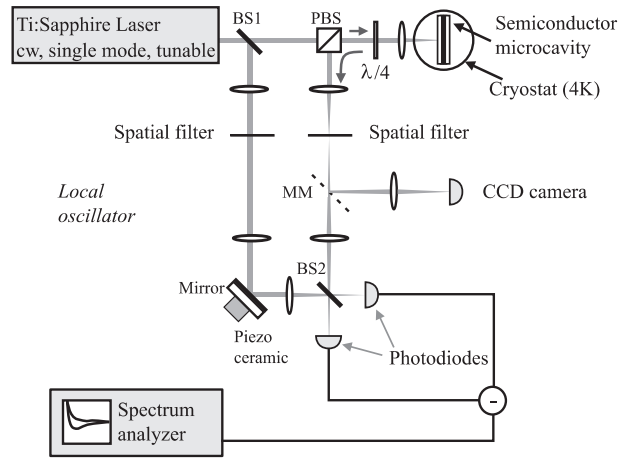}} \caption{Experimental setup. The microcavity sample is
excited using a Ti:Sa laser. The quarter wave plate in front of the sample ensures excitation with a circular
polarization. A spatial filter is placed in the near field of the reflected beam. Using the movable mirror MM, the beam
can be either observed on a CDD camera, again in the near field (which allows to study the spatial effects and to choose
the position of the spatial filter), or sent towards a homodyne detection setup \cite{bachor}. Beam splitter 1 (BS1)
allows to take a part of the excitation beam, used as a local oscillator that is recombined on BS2 with the spatially
filtered reflected beam.} \label{fig2}
\end{figure}

\begin{figure}[h]
\centerline{\includegraphics[clip=,width=8.1cm]{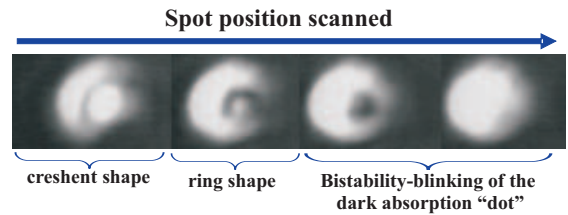}} \caption{Near-field images of the reflected beam, for
different positions of the excitation spot over the sample. The laser intensity is 2mW and the excitation energy is 831.69
nm. The last two images are obtained for the same position ; we observed a blinking between these two states, showing
evidence of bistability. The transition from one bistable state to the other is due to mechanical vibrations.}
\label{fig3}
\end{figure}

The sample used in these experiments is a high finesse GaAs microcavity containing one InGaAs uantum well with low indium
content and is described in more detail in Ref.~\cite{prl}. The Rabi splitting is 2.8 meV. The linewidth (FWHM) of the
lower polariton at 4K is of the order of 200 $\mu$eV. The light source is a single-mode tunable cw Ti:sapphire laser with
a linewidth of the order of 1 MHz. The laser beam is power-stabilized using an electro-optic modulator and spatially
filtered by a 2 m-long single mode fiber. The spot diameter is 50 $\mu$m. In all experiments the lower polariton branch is
excited close to resonance at normal incidence with a $\sigma^{+}$ polarized beam. The outgoing beam is spatially filtered
in order to select a region of the sample of about 10 $\mu$m in diameter (see Fig.~\ref{fig2}) and analyzed by means of a
homodyne detection setup \cite{bachor}. For this the light going out of the microcavity is mixed on two photodiodes with
an auxiliary laser beam, the local oscillator, that has the same frequency as the driving laser. The frequency spectrum of
the beat signal detected by the two photodiodes is analyzed with an RF spectrum analyzer. The reflected laser light and
the light scattered at the laser frequency yield a large interference peak at zero frequency, which is filtered out. The
incoherent luminescence emission yields a broad noise signal. The phase of the local oscillator can be varied, allowing to
explore all the quadrature components present in the emitted noise.

We first studied the near field of the reflected beam. At sufficiently low excitation intensity, the spot shows a dark
vertical line corresponding to absorption occurring on the polariton resonance. The variation of absorption with the
position is due to the slight angle between the cavity mirrors, and the dark line is a line of equal thickness of the
microcavity. At higher laser intensities, even well below the bistability threshold, one observes a strong distortion of
the resonance line as shown in Fig.~\ref{fig3}. When scanning the spot position on the sample (which amounts to scanning
the polariton energy) one can see an evolution from a crescent shape to a ring shape, and then to a dot shape. The shape
of the resonance region can be understood as resulting from compensation between the nonlinear energy shift due to the
intensity variations linked to the gaussian shape of the beam, and the linear energy shift due to the cavity thickness
variations.

According to the above model, the best squeezing rate is expected in the vicinity of a turning point of bistability, as in
other nonlinear systems (see for example \cite{hilico}). Therefore we performed noise measurements close to the turning
points, where the resonance region is dot-shaped. In this case, only the central part of the beam has entered the
microcavity, the remaining part has been reflected by the input mirror. In order to avoid averaging the noise over the
whole reflected beam, we have selected the light emitted by the resonance region (about 10 $\mu$m) by means of a spatial
filter. This means that the combination of geometrical and nonlinear effect leads to selecting a squeezed class of
polariton that has broader dispersion in $k$ than the excitation laser spot.

\begin{figure}[t]
\centerline{\includegraphics[clip=,width=8.1cm]{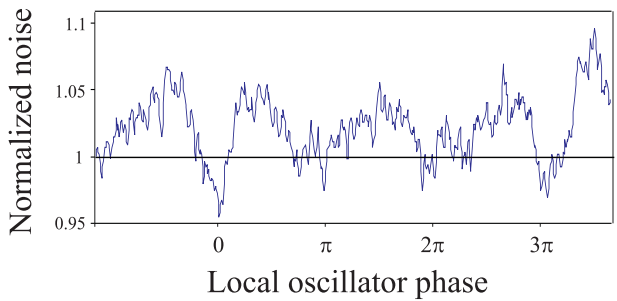}} \caption{Noise signal measured with the homodyne detection
system, normalized to standard quantum noise, when the local oscillator phase is varied allowing to explore all the
quadrature components of the noise. The oscillation period is $\pi$ rather than 2$\pi$ because phases $\phi$ and
$\phi+\pi$ are equivalent in the noise. Rather large variations are seen from one period to the next because of mechanical
vibrations. The cavity-exciton detuning is 0.3 meV and the excitation intensity is 2.2 mW.} \label{fig4}
\end{figure}

The resulting noise signal is shown in Fig.~\ref{fig4}. After corrections for detection efficiency, we measure a squeezing
in the reflected field of 6\%. Squeezing is achievable only in a narrow range of detunings and laser intensities below the
bistability threshold. The value of $\beta$ for the detuning $\delta$=0.3 meV (evaluated at low excitation intensity) is
0.5 $\pm$ 0.2 $\beta_{c}$. The squeezing value predicted by the phenomenological model under these conditions is about
12\%, as can be seen in Fig.~\ref{fig1}. The order of magnitude is in satisfactory agreement with the experimental value,
in view of the approximations made in the model.

From the measurement of the reflected field fluctuations we deduce a squeezing of 9\% of the internal polariton field
using the method of Ref.~\cite{walls}. In contrast to the case of a Kerr medium in a cavity, the squeezing of the
reflected field is smaller than the internal mode squeezing, because the squeezed mode is a polariton mode and not a
photon mode, which causes additional losses at the output of the cavity (since only the photon fraction of the polariton
goes through the cavity mirror). In some way, measuring the output light field is like looking at the polariton mode
through a beamsplitter with an amplitude transmission coefficient equal to $C_{0}$ (the Hopfield coefficient representing
the photon fraction of the polariton) which leads to a reduction of the measured squeezing with respect to the internal
polariton squeezing.

In conclusion, we have reported the first experimental demonstration of squeezed polariton generation in semiconductor
microcavities, using polariton degenerate four-wave mixing. The polariton squeezing generates squeezing of the outgoing
light, that is measured to be $6\%$. The squeezing was obtained just above the bistability threshold, close to a turning
point, and the bistable region was selected by spatial filtering. A squeezing of $9\%$ of the part-light, part-matter
polariton field can be inferred from the measurements. This result opens the way to the generation of nonclassical states
of light and matter in semiconductor microcavities. Such nonclassical effects should also be observed in nondegenerate
four-wave mixing, since quantum correlations between the polariton signal and idler fields have been recently predicted
\cite{quattropani}.

\end{document}